\documentclass[aps,prl,reprint,superscriptaddress,twocolumn,floatfix]{revtex4-2}
\usepackage{amsmath,amssymb}
\usepackage{bm}
\usepackage{xcolor}
\usepackage{hyperref}
\usepackage{graphicx}
\usepackage{lineno}
\newcommand{\snn}{\sqrt{s_{NN}}}
\newcommand{\Npart}{\langle N_{\rm part}\rangle}
\newcommand{\dQ}{\Delta Q}
\newcommand{\Bavg}{\langle B\rangle}
\newcommand{\RBQ}{\Bavg/\dQ\times\Delta Z/A}
\newcommand{\Pu}{P_{u}}

\begin{document}

\title{Quantitative Junction Transport Modeling of RHIC Charge and Baryon Stopping}

\author{Ashish Pandav}
\email{ashishp.pandav@gmail.com}
\affiliation{Center for Nuclear Research, Department of Physics, Kent State University, Kent, Ohio 44242, USA}

\author{Zhangbu Xu}
\email{zxu22@kent.edu}
\affiliation{Center for Nuclear Research, Department of Physics, Kent State University, Kent, Ohio 44242, USA}

\author{Bedangadas Mohanty}
\email{bedanga@niser.ac.in}
\affiliation{School of Physical Sciences, National Institute of Science Education and Research, HBNI, Jatni, Odisha 752050, India}

\begin{abstract}
Charge and baryon stopping in Ru+Ru and Zr+Zr isobar collisions are modeled by HIJING with junction. With realistic nuclear geometry and a tuned diquark splitting, it reproduces STAR's net-baryon yield, charge difference, and scaled baryon-to-charge ratio at midrapidity. Junctions set the ratio's magnitude while neutron skins drive its centrality trend. At the LHC, large asymmetric pair production in HIJING obscures junction effects, causing apparent tension with data. This tuned generator reproduces RHIC data and predicts O+O behavior, setting a quantitative benchmark for baryon transport.
\end{abstract}

\maketitle
Baryon number is conserved, and in the quark model each valence quark carries one third of it. An alternative picture was proposed early in the development of QCD: the three valence quarks of a baryon are connected by a Y-shaped configuration of gluon fields, the baryon junction, and it is this gluonic object that traces the baryon number~\cite{Artru:1974zn,Rossi:1977cy}. Because the junction is made of soft gluons it carries a small fraction of the nucleon momentum and can be stopped near midrapidity in a high-energy collision, while the valence quarks continue toward beam rapidity and fragment into mesons~\cite{Kharzeev:1996sq}. Regge theory then predicts a net-baryon rapidity density that falls as $\exp(-\alpha_B\,\delta y)$ with the rapidity shift $\delta y$ from the beam, with $\alpha_B$ between 0.42 and 1~\cite{Kharzeev:1996sq,Lewis:2022arg}. Diquark transport in string models, where the baryon number stays with the valence quarks, gives a much steeper fall~\cite{Lewis:2022arg}. This distinction has motivated renewed interest in baryon stopping and baryon-transport mechanisms~\cite{Lv:2023fxk,Magdy:2025udq,Pratt:2023pee,Magdy:2024dpm,Garcia-Montero:2024jev,Pihan:2026leq}.

Recent RHIC measurements~\cite{STAR:2024lvy} further strengthen the case for junction-mediated transport. 
The net-proton yield at midrapidity in Au+Au collisions from $\sqrt{s_{NN}}=7.7$ to $200\,\text{GeV}$ exhibits an exponential dependence on rapidity gap with slope $\alpha_B = 0.64 \pm 0.05$, independent of centrality~\cite{Lewis:2022arg,STAR:2024lvy}. Net-$\Lambda$, net-$\Xi$, and net-$\Omega$ yields exhibit the same slope within uncertainties, consistent with a flavor-blind baryon carrier~\cite{Tsang:2024zsq}. The isobar collisions $^{96}$Ru+$^{96}$Ru and $^{96}$Zr+$^{96}$Zr provide the most direct test: the nuclei have the same mass number but differ in charge by four units. If the valence quarks that transport electric charge also transported baryon number, the scaled ratio $\RBQ$ would be expected to be near unity. Instead, STAR reports a value of $1.84\pm0.02\,({\rm stat})\pm0.09\,({\rm syst})\pm0.16\,({\rm feed\mbox{-}down})$ in central collisions, decreasing toward unity in peripheral collisions~\cite{STAR:2024lvy}. This strong excess above unity indicates significant baryon transport that is not correlated with the transported charge.

No event generator without junctions reproduces this observable. UrQMD~\cite{Bass:1998ca,Bleicher:1999xi} and AMPT~\cite{Lin:2004en} yield ratios between 0.5 and 0.7, even below the naive expectation of unity, because these models transport more charge than baryon number to midrapidity, with the excess appearing in the mesonic sector~\cite{Lewis:2022arg,STAR:2024lvy,Ross:2025qxr}. The TRENTO initial-state model reproduces the qualitative centrality dependence through the larger neutron skin of $^{96}$Zr~\cite{Moreland:2014oya,Li:2019kkh,Xu:2021qjw,Pihan:2026leq}, which reduces the participating proton fraction in peripheral Zr+Zr collisions, but cannot account for the magnitude of the observed ratio due to the absence of dynamical baryon transport.

\textsc{HIJING}/B$\overline{\text{B}}$ is the only Monte Carlo event generator in which baryon junctions are embedded in the incoming nucleons and participate directly in the scattering~\cite{Vance:1998vh,Vance:1999pr,ToporPop:2004lve,ToporPop:2005grb}. It was originally developed to describe baryon stopping at SPS and RHIC energies but had not been confronted with the RHIC isobar measurement. In this work we use version~2.0 of the model, incorporating realistic proton and neutron density distributions for the isobars, tuning several key physics parameters, to compare with STAR’s charge and baryon stopping results. We examine the effect of switching the junction mechanism on and off and compare the model with Au+Au and $p$+$p$ data at $\sqrt{s_{NN}}=200\,\text{GeV}$, and provide predictions for O+O collisions. We caution that the model is not extended to LHC energies and beyond because a different mechanism, rapidity-asymmetric baryon–antibaryon pair production in HIJING, contributes significantly to the midrapidity net-baryon signal at these ultra-high energies and obscures any junction contribution. 

The simulations use event generator \textsc{HIJING}/B$\bar{B}$~v2.0, which is built on \textsc{HIJING}~1.382~\cite{Wang:1991hta}. Soft interactions are modeled through excitation of quark--diquark strings of wounded nucleons, while hard scatterings are treated via minijet production with Glauber geometry, nuclear shadowing, and jet quenching as in the baseline model. In the default configuration, the diquark remains near beam rapidity and carries the baryon number, whereas the quark is pulled toward midrapidity, transporting electric charge into the meson sector.

Two additional mechanisms are implemented. First, baryon junction exchange~\cite{Vance:1998vh,Kharzeev:1996sq} is enabled: in a nucleon--nucleon collision the junction may be exchanged with cross section $\sigma_J$, after which the nucleon fragments into three strings meeting at a Y-shaped junction. We use $\sigma_J=28$~mb, which reproduces the net-baryon yield in the isobars. A baryon is formed from the junction and three sea quarks, and its rapidity is sampled from $\exp(\lambda_J y)$ between the outermost string ends, carrying baryon number without transporting charge. The value $\lambda_J=0.55$ is chosen to match the measured rapidity slope of the net-proton distribution in Au+Au collisions. Second, strong color fields~\cite{ToporPop:2005grb,ToporPop:2006jqg} increase the effective string tension above the Lund value $\kappa_0=1$~GeV/fm. Following Refs.~\cite{ToporPop:2005grb,ToporPop:2006jqg}, we use $\kappa=2\kappa_0$ for nucleus--nucleus and $\kappa=1.5\kappa_0$ for $p$+$p$ collisions. Junction exchange and enhanced string tension are the only physics additions in v2.0. As discussed in Ref.~\cite{ToporPop:2004lve}, this version does not include explicit junction--antijunction loop mechanisms; pair production arises solely from fragmentation at the enhanced tension. Transverse-momentum kicks to stopped junction baryons and to fragmentation diquarks are disabled, since neither option produces a $p_T$ dependence of $\bar p/p$ consistent with the LHC measurements~\cite{ALICE:2010hjm}, while the net quantities and the isobar observables, integrated over $p_T$, are unchanged within a few percent.

The quark--diquark splitting probability $P_u$ controls whether the extracted valence quark from a wounded proton is a $u$ quark (probability $P_u$) or a $d$ quark ($1-P_u$), with the isospin-mirrored choice for a neutron. The default SU(6) value is $P_u=2/3$. Since the lone quark fragments near midrapidity, $P_u$ determines the mean valence charge transported and therefore the isobar charge difference. We treat $P_u$ as an effective string parameter and use $P_u=0.58$. This setting matches the charge-difference ($\Delta Q$) observable while the midrapidity hadron yields are insenstive to its variation and stay within $0.3\%$. 

Unless stated otherwise, all other settings follow the \textsc{HIJING}/B$\bar{B}$~v2.0 defaults. Three parameters are adjusted in this work:
$\sigma_J = 28$~mb, $\lambda_J = 0.55$, and $P_u = 0.58$. The junction-off reference sample uses the same parameters but with junction exchange disabled, isolating the junction contribution.


The proton and neutron densities of $^{96}$Ru and $^{96}$Zr are sampled independently from Woods--Saxon distributions,
\[
\rho(r) \propto \left[1+\exp\!\left(\frac{r-R}{a}\right)\right]^{-1},
\]
with species-dependent half-density radii $R$, diffuseness parameters $a$, and quadrupole and octupole deformations $(\beta_2,\beta_3)$ taken from Ref.~\cite{Nijs:2021kvn} (case~5). These parameters reproduce the STAR multiplicity and flow constraints~\cite{STAR:2021mii}. For $^{96}$Ru we use $R_p=5.053$~fm, $a_p=0.48$~fm, $R_n=5.073$~fm, $a_n=0.49$~fm, and $\beta_2=0.154$. For $^{96}$Zr we use $R_p=4.912$~fm, $a_p=0.508$~fm, $R_n=5.007$~fm, $a_n=0.564$~fm, $\beta_2=0.062$, and $\beta_3=0.202$. The resulting neutron skin of Zr is of halo character, $\Delta r_{np}\approx 0.16$~fm, whereas Ru exhibits nearly overlapping proton and neutron densities, $\Delta r_{np}\approx 0.03$~fm. 

The default \textsc{HIJING} initialization sorts nucleons along the beam direction and assigns the first $Z$ nucleons as protons, correlating charge with spatial order. As noted in Ref.~\cite{Ross:2025qxr}, this artefact is removed by assigning nucleon species prior to spatial sampling. For $^{197}$Au, the charge distribution of Ref.~\cite{DeVries:1987atn} is used together with neutron diffuseness $a_n=0.641$~fm, corresponding to $\Delta r_{np}=0.16$~fm. For $^{16}$O we adopt the standard three-parameter Fermi form of \textsc{HIJING} with identical proton and neutron profiles. All nucleus--nucleus samples are generated minimum-bias across the full impact-parameter range; the $p{+}p$ sample is inelastic. Table~\ref{tab:prod} summarizes the event samples.

\begin{table}[tb]
\caption{Event samples at $\sqrt{s_{NN}}=200$~GeV.}
\label{tab:prod}
\begin{tabular}{llll}
\hline
system & junction & geometry & events \\
\hline
Ru+Ru, Zr+Zr & on  & skin, deformation & $60\times10^6$ \\
Ru+Ru, Zr+Zr & off & skin, deformation & $30\times10^6$ \\
Ru+Ru, Zr+Zr & on, $P_u=2/3$ & skin, deformation & $30\times10^6$ \\
Au+Au & on & skin & $30\times10^6$ \\
O+O & on & spherical & $200\times10^6$ \\
$p$+$p$ & on & -- & $1\times10^6$ \\
\hline
\end{tabular}
\end{table}


We adopt the STAR definitions~\cite{STAR:2024lvy} for the midrapidity net charge and net baryon number with all yields taken at $|y|<0.5$ and integrated over $p_T$. As in the STAR analysis, weak-decay products are included. The charge difference between the isobars is obtained from the double-ratio estimator~\cite{Lewis:2022arg},
\begin{equation}
\Delta Q \simeq
N_{\pi}(R_{2\pi}-1)
+ N_{K}(R_{2K}-1)
+ N_{p}(R_{2p}-1),
\label{eq:dq}
\end{equation}
where $R_{2\pi} \equiv (N_{\pi^+}/N_{\pi^-})_{\rm Ru}/(N_{\pi^+}/N_{\pi^-})_{\rm Zr}$, and $N_\pi$ is the yield averaged over both charges and both systems; analogous definitions apply to kaons and protons. Equation~(\ref{eq:dq}) is the estimator used by STAR and removes the small overall multiplicity difference between Ru+Ru and Zr+Zr. The average midrapidity net-baryon number is
\[
\langle B\rangle = \tfrac{1}{2}\big(B_{\rm Ru+Ru} + B_{\rm Zr+Zr}\big),
\]
and these quantities define the scaled ratio
\[
R_{B/Q} = \frac{\langle B\rangle}{\Delta Q} \times \frac{\Delta Z}{A},
\]
with the bulk isospin difference $\Delta Z/A = 4/96$. Centrality is determined from the charged-particle multiplicity in $|\eta|<0.5$, following the STAR procedure. Statistical uncertainties are evaluated using the delta theorem~\cite{Casella:2002,Luo:2011tp}.

\begin{figure}[!htbp]
\includegraphics[width=\columnwidth]{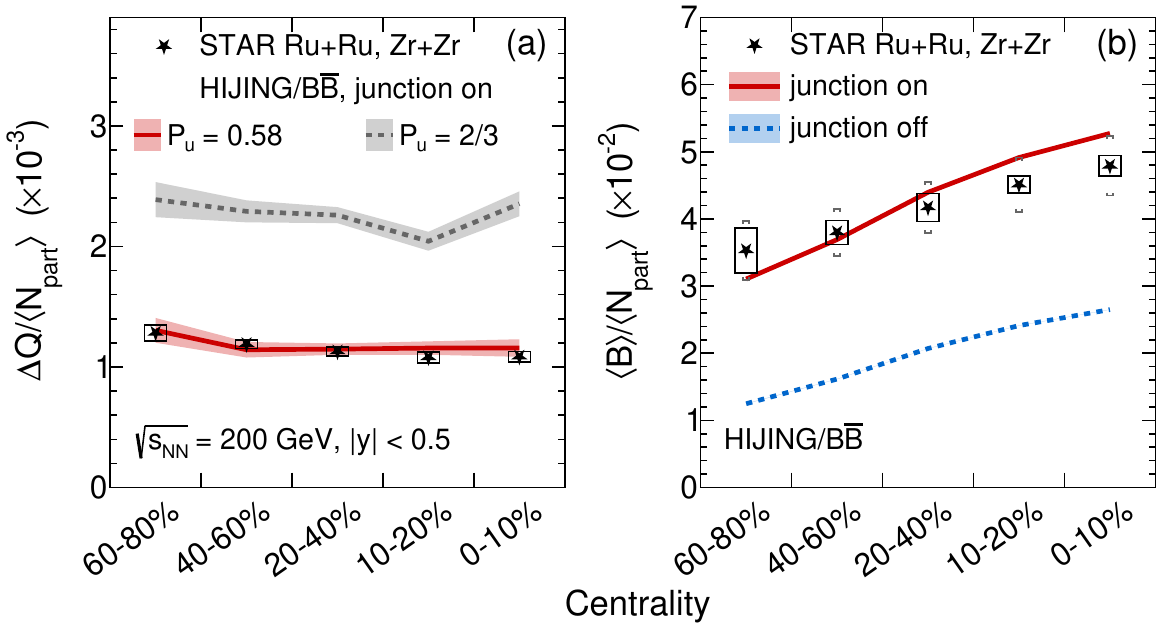}

\caption{
(a) Net-charge difference per participant between Ru+Ru and Zr+Zr collisions at $\sqrt{s_{NN}}=200$~GeV and $|y|<0.5$ as a function of centrality: STAR data~\cite{STAR:2024lvy} (bars: statistical; boxes: systematic uncertainties) compared with \textsc{HIJING}/B$\bar{\rm B}$ for $P_u=0.58$ and $P_u=2/3$; bands are the model statistical uncertainties. (b) Net-baryon number per participant, averaged over Ru+Ru and Zr+Zr, with the junction on and off; STAR brackets show the feed-down uncertainty of the neutron estimate. Model uncertainties are smaller than the line width.
}
\label{fig:dq}
\end{figure}

Figure~\ref{fig:dq}(a) shows $\Delta Q/\langle N_{\rm part}\rangle$ as a function of centrality. With the default quark--diquark splitting probability $P_u = 2/3$, the junction-on calculation yields $(2.0$--$2.4)\times10^{-3}$, roughly twice the measured $(1.1$--$1.3)\times10^{-3}$. The excess arises entirely from the charge sector: too much of the valence charge of the additional Ru protons is transported to midrapidity. Using $P_u = 0.58$ brings the model into agreement with the STAR data in all centrality classes, giving $\sim1.16\times10^{-3}$ in 0--10\% and $\sim1.30\times10^{-3}$ in 60--80\%. This value of $P_u$ is applied throughout the analysis. Adjusting $P_u$ modifies the $\pi$, $K$, $p$, and $\bar p$ yields, and the midrapidity net-baryon number, by less than $0.3\%$ (see Supplemental Material, Note~S1); the junction sector is unaffected. 

With the junction mechanism switched off, $\Delta Q$ remains essentially unchanged, $(1.0$--$1.4)\times10^{-3}$ versus $(1.1$--$1.3)\times10^{-3}$ with it on, consistent within statistical uncertainties across all centralities. In contrast, the midrapidity net-baryon number is reduced by roughly a factor of two (Fig.~\ref{fig:dq}(b)), demonstrating that junction exchange contributes to baryon transport but does not enter the charge-difference observable.

Figure~\ref{fig:dq}(b) shows $\Bavg/\Npart$. With the junction on, and $\sigma_J$ chosen to match the level of the measured yield, the model is consistent with the data within their uncertainties in all five centrality classes. It rises somewhat more steeply with centrality than the data. With the junction off the net-baryon number is a factor two lower, $\sim1.2$ to $2.7\times10^{-2}$, and far below the data everywhere: the junction supplies about half of the midrapidity net-baryon number. Baryon stopping in the diquark strings alone is insufficient at this energy, as found at the SPS when the junction was introduced~\cite{Vance:1998vh}.

\begin{figure}[!htbp]
\includegraphics[width=\columnwidth]{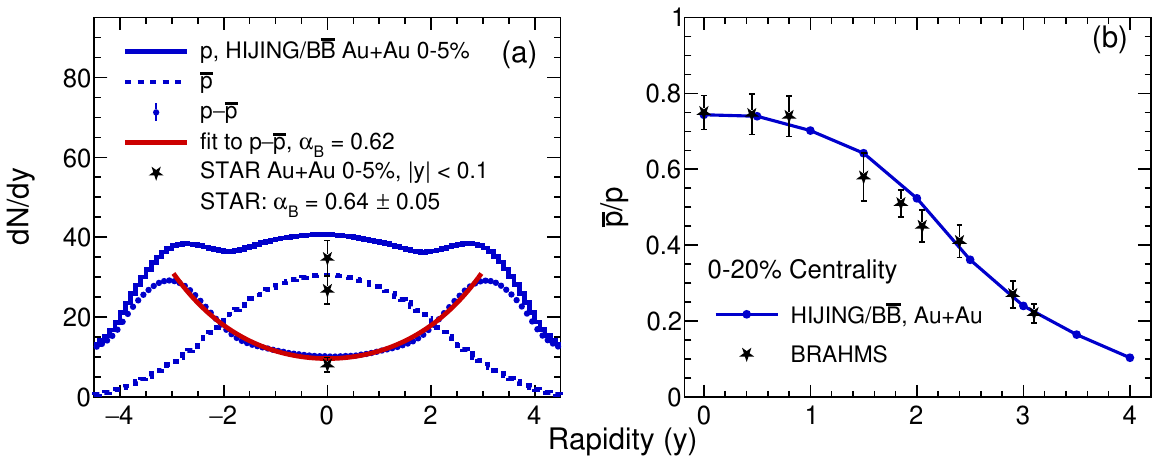}
\caption{HIJING/B$\bar{\rm B}$ with the junction on compared with RHIC data at $\snn=200$~GeV. (a) Proton, antiproton and net-proton rapidity densities in 0--5\% Au+Au collisions, the fit to the net proton described in the text, and the STAR midrapidity yields~\cite{STAR:2008med} at $|y|<0.1$. (b) $\bar p/p$ versus rapidity in 0--20\% collisions with the BRAHMS data for the same centrality class~\cite{BRAHMS:2002efn}. }
\label{fig:rhic}
\end{figure}

The same parameter set is compared with the Au+Au data at $\snn=200$~GeV in Fig.~\ref{fig:rhic}. The model yields include weak-decay products throughout, as do the STAR yields and the BRAHMS ratios. Panel (a) shows the proton, antiproton and net-proton rapidity densities in 0--5\% Au+Au collisions with the STAR midrapidity yields~\cite{STAR:2008med}. The model gives $dN/dy=40.5$ for protons and $30.3$ for antiprotons at $|y|<0.5$, with statistical uncertainties below 0.1\% here and throughout, against $34.7\pm4.4$ and $26.7\pm3.4$; the net-proton yield is $10.2$ against $8.0\pm1.8$. STAR extracts $\alpha_B$ from the midrapidity net-proton yield, so the fit is made to it: the form $C[e^{-\alpha_B(y_b-y)}+e^{-\alpha_B(y_b+y)}]$ with $y_b=5.36$, fitted over $|y|<3$, the range before the distribution turns over toward beam rapidity, gives $\alpha_B=0.62$; STAR obtains $0.64\pm0.05$ from the midrapidity yield as a function of beam energy, where the rapidity gap $\delta y=y_b$ is varied through $\sqrt{s_{NN}}$~\cite{STAR:2024lvy}. The charged-particle density is $dN_{\rm ch}/d\eta=591$ against $691\pm49$~\cite{STAR:2008med}; the difference is in the produced pairs, which cancel in the net quantities of Fig.~\ref{fig:dq}. Panel (b) shows $\bar p/p$ as a function of rapidity in 0--20\% collisions, the centrality class of the BRAHMS measurement~\cite{BRAHMS:2002efn}; the model follows the measured decrease from $0.75\pm0.04$ at $y=0$ to $0.22\pm0.03$ at $y=3.1$. In $p$+$p$ collisions at the same energy the model gives an inclusive net-proton density of 0.032 per unit rapidity per non-single-diffractive event against the STAR value of $0.025\pm0.004$~\cite{STAR:2008med}, and $\bar p/p=0.73$ against $0.78\pm0.07$ from BRAHMS~\cite{BRAHMS:2004gnf} and $0.82\pm0.05$ from STAR~\cite{STAR:2008med}.

The model with the junction reproduces the Au+Au and $p$+$p$ baryon sector at the level of the experimental uncertainties. Of the quantities in Fig.~\ref{fig:rhic}, only the net-proton rapidity slope was used to fix a parameter, the sampling exponent.

ALICE measures $\bar p/p=0.957\pm0.015$ at midrapidity in $p$+$p$ collisions at $\sqrt s=0.9$~TeV, with the weak-decay feed-down subtracted~\cite{ALICE:2010hjm}. We ran the model at this energy with the parameters above, weak decays off, and the $p$+$p$ string tension $\kappa/\kappa_0=1.5$ used at 200~GeV~\cite{ToporPop:2006jqg}. The model gives $\bar p/p=0.87$, and the net-proton yield is more than twice the measurement. Raising the string tension to 2.25, the energy-dependent value prescribed for this energy by the model's authors~\cite{ToporPop:2010qz}, doubles the net-proton excess and leaves $\bar p/p$ at 0.88. The earlier version of the model overshoots the measured proton--antiproton asymmetry by about 45\%~\cite{ToporPop:2010qz}, and ALICE lists HIJING/B at $\bar p/p=0.83$~\cite{ALICE:2010hjm}. To locate the excess, the junction baryons were identified by their origin in the event record and set aside, and every antibaryon was matched to the nearest of the other baryons in rapidity. Baryon number conservation leaves some of these other baryons unmatched: they are the ones transported by the diquark strings of nucleons whose junction was not exchanged. A matched pair whose baryon falls inside the rapidity window and whose antibaryon falls outside contributes a net baryon that was never transported, the pair asymmetry.  Table~\ref{tab:lhc} gives the three components of the net-proton yield at $|y|<0.5$ at $\sqrt s=200$ and 900~GeV.

\begin{table}[tb]
\caption{Components of the midrapidity net-proton yield, $dN_{p-\bar p}/dy$ at $|y|<0.5$, in $p$+$p$ collisions at $\sqrt s=200$ and 900~GeV: HIJING/B$\bar{\rm B}$ with the junction on, primary particles per inelastic event, decomposed as described in the text. Statistical uncertainties are $\sim$0.0001 for the junction and diquark components and $\sim$0.0003 for the pair asymmetry. Data: STAR at 200~GeV~\cite{STAR:2008med}, converted to the model convention with the model ratios 0.56 (primary to inclusive) and 0.72 (inelastic to non-single-diffractive); ALICE at 900~GeV as $p\,(1-\bar p/p)$ from Refs.~\cite{ALICE:2011gmo,ALICE:2010hjm}. Data uncertainties are total.}
\label{tab:lhc}
\small
\setlength{\tabcolsep}{3pt}
\begin{tabular}{lccc}
\hline
component & 200 GeV & 900 GeV & ratio \\
\hline
junction & 0.0115 & 0.0043 & 0.37 \\
diquark strings & 0.0018 & 0.0010 & 0.56 \\
pair asymmetry & 0.0003 & 0.0041 & 16 \\
net $p$, model & 0.0135 & 0.0094 & 0.70 \\
data (net $p$) & $0.0101\pm0.0016$ & $0.0036\pm0.0013$ & $0.36\pm0.14$ \\
\hline
\end{tabular}
\end{table}

The junction contribution falls by a factor of 0.37 between 200 and 900~GeV, consistent with the measured reduction, $0.36\pm0.14$. At 900~GeV, the junction component alone ($0.0043$) accounts for the observed net-proton yield, $0.0036\pm0.0013$, implying that the excess arises from the pair asymmetry. When a diquark--antidiquark pair is produced adjacent to the valence quark at the slow end of a string, color ordering causes the baryon to form at that end and the antibaryon one hadron farther toward beam rapidity. The resulting rapidity separation of about one unit places the baryon preferentially closer to midrapidity~\cite{Andersson:1983ia}. Because this contribution scales with pair production, it grows from about $2\%$ of the net proton at 200~GeV to a value comparable to the entire measured net proton at 900~GeV, and increases further with the string tension. The 0.9~TeV comparison therefore constrains HIJING string fragmentation at LHC multiplicities rather than baryon transport, and we foresee future developments addressing this issue before the junction version of HIJING can be used at LHC energies. In Au+Au collisions at 200~GeV, the heaviest system studied here, the same contribution is below $\sim2\%$ of the net proton, as in $p$+$p$ at this energy.

\begin{figure}[tb]
\includegraphics[width=\columnwidth]{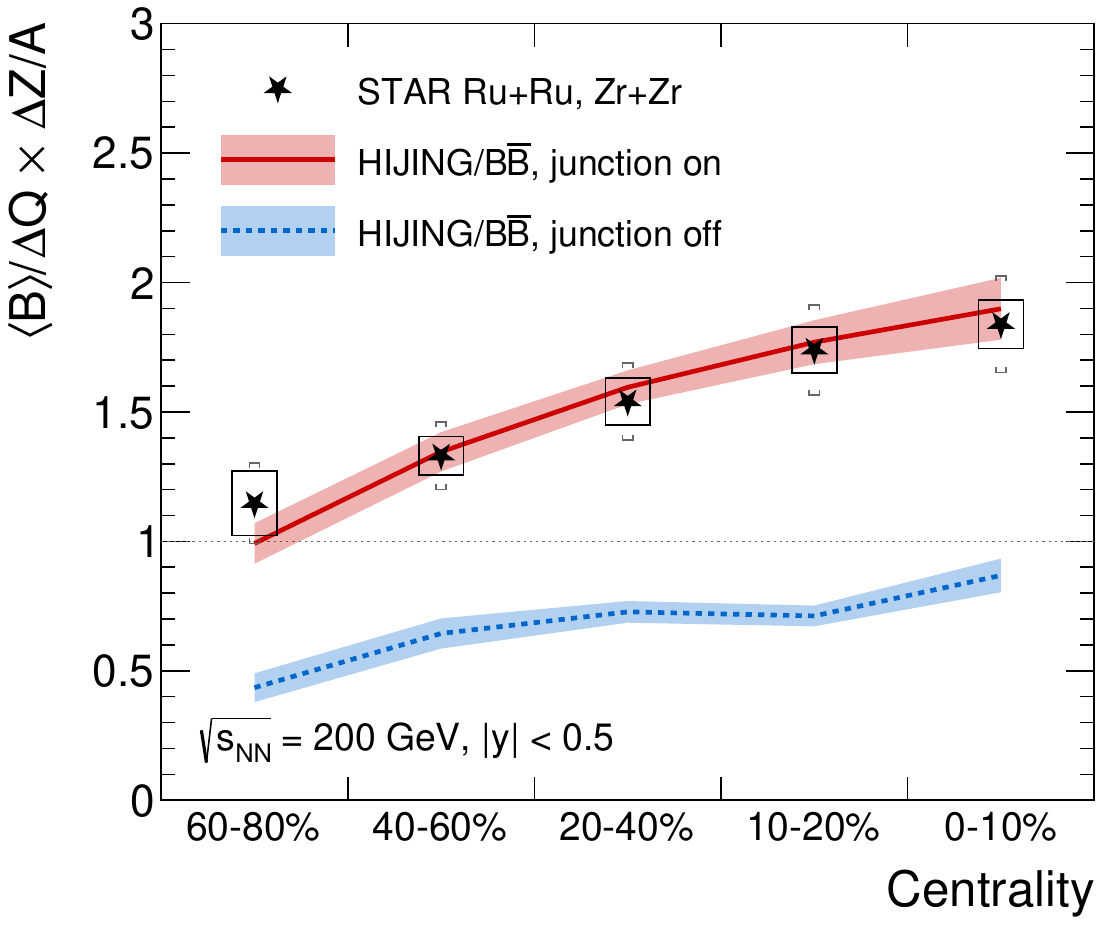}
\caption{$\RBQ$ at $|y|<0.5$ in Ru+Ru and Zr+Zr collisions at $\sqrt{s_{NN}}=200$~GeV versus centrality. STAR data~\cite{STAR:2024lvy} (bars: statistical, boxes: systematic, brackets: feed-down uncertainty of the neutron estimate) compared with HIJING/B$\bar{\rm B}$ with the junction on and off; bands are the model statistical uncertainties.}
\label{fig:ratio}
\end{figure}

Figure~\ref{fig:ratio} shows $\RBQ$. Since $\dQ$ and $\Bavg$ are both described, the ratio description follows. The agreement is within the uncertainties in all classes. With the junction off the ratio is $\sim$0.44 to 0.87 (Fig.~\ref{fig:ratio}), in the range of the other models without junctions~\cite{STAR:2024lvy}. The two pieces of the ratio have different origins. Its magnitude comes from the junction: the transported baryon is built from sea quarks and carries no memory of the valence charge, so it adds to $B$ and not to $\dQ$. Its centrality dependence comes from the nuclear geometry: the neutron skins lower the fraction of participating protons in peripheral collisions, far more in Zr+Zr than in Ru+Ru, which raises $\dQ$ there. When the skin is removed the ratio becomes flat within errors (Note~S2 of the Supplemental Material), while the magnitude stays near $\sim$1.5. When the deformation is removed nothing changes.

\begin{figure}[tb]
\includegraphics[width=\columnwidth]{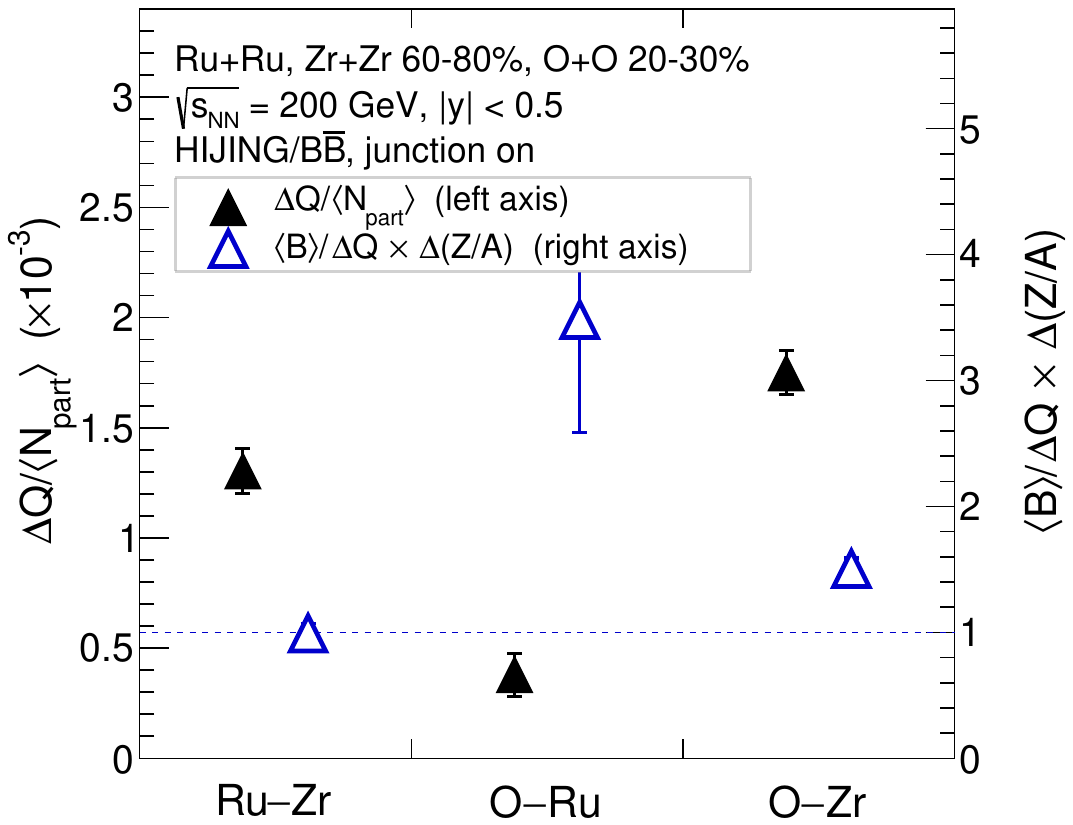}
\caption{O+O compared with the isobars (Ru+Ru, Zr+Zr) at the same $\Npart$, the O+O 20--30\% and the isobar 60--80\% centrality classes at $\sqrt{s_{NN}}=200$~GeV, HIJING/B$\bar{\rm B}$ with the junction on: $\dQ/\Npart$ and $\Bavg/\dQ\times\Delta(Z/A)$ for the pairs Ru$-$Zr, O$-$Ru and O$-$Zr. Uncertainties are statistical.}
\label{fig:oo}
\end{figure}

STAR has recorded O+O collisions at $\snn=200$~GeV. $^{16}$O has $Z/A=1/2$ and no neutron skin, so it provides a reference with a known participant isospin. Since the net-baryon number per participant grows with centrality, the comparison with the isobars has to be made at the same $\Npart$. We compare O+O collisions with peripheral Ru+Ru and Zr+Zr collisions at the same average number of participants: the isobar 60--80\% class has $\Npart=14.4$ (Ru) and $13.9$ (Zr), and the O+O 20--30\% class has $\Npart=14.0$. Three differences can be taken between the three systems, Ru+Ru minus Zr+Zr, O+O minus Ru+Ru and O+O minus Zr+Zr, written Ru$-$Zr, O$-$Ru and O$-$Zr, with bulk isospin levers $\Delta(Z/A)=0.042$, $0.042$ and $0.083$, respectively. Figure~\ref{fig:oo} shows $\dQ/\Npart$ and $\Bavg/\dQ\times\Delta(Z/A)$ for the three pairs.

The net-baryon number per participant is the same in the three systems to within 2\%, so the differences between the pairs are entirely in the charge difference. The Ru$-$Zr pair gives $\dQ/\Npart=(1.30\pm0.10)\times10^{-3}$ and a scaled ratio $\RBQ=0.99\pm0.08$. The O$-$Zr pair has twice the lever but only $1.35$ times the charge difference, $(1.75\pm0.10)\times10^{-3}$, and a ratio of $1.51\pm0.09$.  Both pairs contain Zr, whose neutron skin removes participating protons in peripheral collisions and adds to the charge difference beyond what the bulk lever gives. The addition is the same in both pairs and weighs more against the smaller lever of Ru$-$Zr. The O$-$Ru pair has the same bulk lever as Ru$-$Zr but no Zr skin, so its charge difference is small, $(0.38\pm0.10)\times10^{-3}$, and the ratio, $3.5\pm0.9$, has a correspondingly larger uncertainty. The O$-$Zr pair, with the largest lever, is the best determined of the three. The model thus predicts the ordering of the pairs: in the charge difference, $\mathrm{O{-}Ru}<\mathrm{Ru{-}Zr}<\mathrm{O{-}Zr}$, and in the scaled ratio, $\mathrm{Ru{-}Zr}<\mathrm{O{-}Zr}<\mathrm{O{-}Ru}$.

In summary, HIJING/B$\bar{\rm B}$ with the baryon junction reproduces the STAR isobar measurement of $\RBQ$ at $\snn=200$~GeV in all centrality classes, with the net-baryon number and the charge difference described separately, and describes baryon stopping in Au+Au and $p$+$p$ collisions at the same energy. The junction sets the magnitude of the ratio, with the junction off the model gives 0.4 to 0.9 like the models without junctions, and the neutron skins of the isobars set its centrality dependence. For O+O collisions at the same $\Npart$ as peripheral isobar collisions, the net-baryon number per participant, the charge difference and the scaled ratio are obtained for the Ru$-$Zr, O$-$Ru and O$-$Zr pairs; the
latter two are ordered $\mathrm{O{-}Ru}<\mathrm{Ru{-}Zr}<\mathrm{O{-}Zr}$ and
$\mathrm{Ru{-}Zr}<\mathrm{O{-}Zr}<\mathrm{O{-}Ru}$, and the STAR O+O data can test
these predictions.

\begin{acknowledgments}
We thank Xin-Nian Wang for providing the HIJING code and for his help in setting it up on GitHub, and the CNR group at Kent State University for stimulating discussions. This work was supported in part by the Office of Nuclear Physics within the U.S. Department of Energy Office of Science under Contract DE-FG02-89ER40531 and DE-SC0012704. B. M. acknowledges support from the ANRF J. C. Bose Fellowship.
\end{acknowledgments}

\bibliographystyle{apsrev4-2}
\bibliography{refs_PRL}

\clearpage
\onecolumngrid
\setcounter{figure}{0}
\setcounter{table}{0}
\renewcommand{\thefigure}{S\arabic{figure}}
\renewcommand{\thetable}{S\arabic{table}}
\newcommand{\hij}{HIJING/B$\bar{\rm B}$}
\begin{center}
{\large\bfseries Supplemental Material: Quantitative Junction Transport Modeling of RHIC Charge and Baryon Stopping}
\end{center}

\subsection{Note S1: Effect of the quark--diquark splitting probability on other observables}
\label{app:pu}
\begin{figure}[!htbp]
    \centering
    \includegraphics[width=0.85\linewidth]{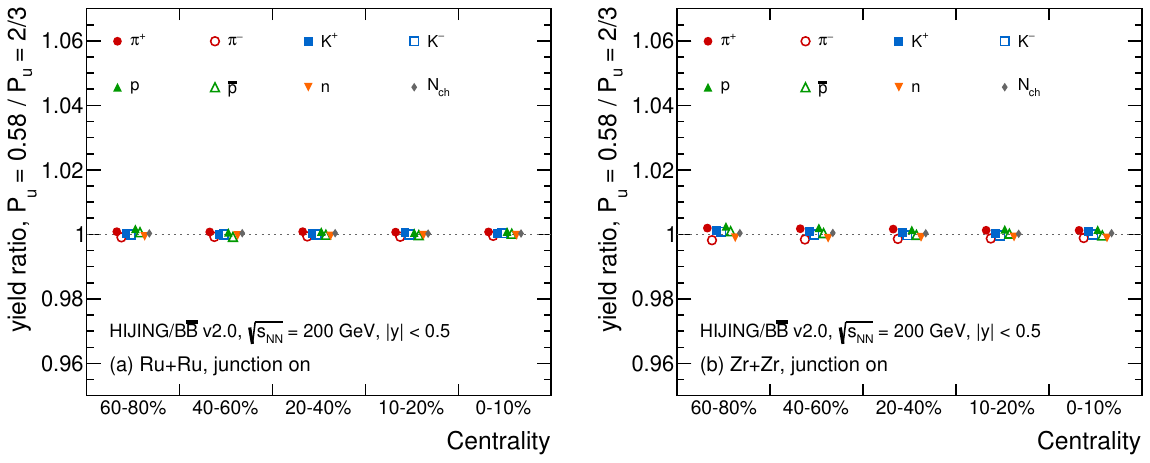}
    \caption{Ratio of per-event midrapidity yields ($|y|<0.5$) obtained with $\Pu=0.58$ to those with $\Pu=2/3$, \hij\ with the junction on, as a function of centrality in (a) Ru+Ru and (b) Zr+Zr collisions at $\snn=200$~GeV. Statistical uncertainties are smaller than the markers.}
    \label{fig:pu}
\end{figure}

Figure~\ref{fig:pu} shows the ratio of the per-event midrapidity yields with $\Pu=0.58$ to those with $\Pu=2/3$, junction on, in Ru+Ru and Zr+Zr collisions for each centrality class. All yields, $\pi^\pm$, $K^\pm$, $p$, $\bar p$, $n$ and the charged multiplicity, change by less than 0.3\%, and the net-baryon number and $\Npart$ in each class by less than 0.3\% and 0.05\%.
The $\dQ$ of Fig.~1(a) of the main text changes by a factor two.
The parameter therefore acts on the isotope charge difference and leaves the measured yields unchanged at the level of the experimental uncertainties.

\subsection{Note S2: Nuclear-structure controls}
\label{app:skin}
\begin{figure}[!htbp]
    \centering
    \includegraphics[width=\linewidth]{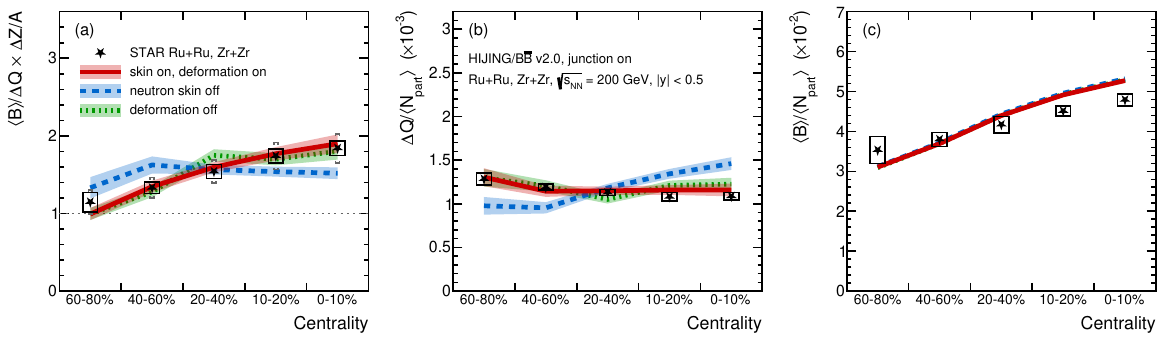}
    \caption{(a) $\RBQ$, (b) $\dQ/\Npart$ and (c) $\Bavg/\Npart$ as a function of centrality in Ru+Ru and Zr+Zr collisions at $\snn=200$~GeV, \hij\ with the junction on: full geometry as in the main text, neutron skin removed, and deformation removed. Bands are statistical uncertainties. STAR data~\cite{STAR:2024lvy} as in Figs.~1 and 3 of the main text, with the same uncertainty elements.}
    \label{fig:skin}
\end{figure}

Figure~\ref{fig:skin} repeats Figs.~1 and 3 of the main text with the neutron skin removed (deformation kept) and with the deformation removed (skin kept).
The net-baryon number per participant is unchanged in both cases. Without the skin the participant proton fraction of Zr is at its bulk value in all classes. The charge difference then falls in peripheral collisions, from $1.3$ to $1.0\times10^{-3}$ in 60--80\%, where the neutron-rich surface of Zr no longer depletes the participating protons, and rises in central collisions, from $1.2$ to $1.5\times10^{-3}$ in 0--10\%, where the diffuse neutron tail of the halo-type skin no longer leaves the participants slightly proton-rich. The ratio $\RBQ$ is then $\sim$1.4 to 1.6, flat within uncertainties.
The centrality dependence of the ratio in Fig.~3 of the main text is therefore produced by the skin, as the TRENTO comparison of Ref.~\cite{STAR:2024lvy} suggested, while its level above unity is not.
Removing the deformation changes neither $\dQ$ nor the scaled ratio beyond the statistical uncertainties.

\end{document}